\documentclass[sigconf]{acmart}
\renewcommand\footnotetextcopyrightpermission[1]{}
\usepackage{algorithm,algorithmicx,algpseudocode}
\usepackage{tikz}
\usepackage{pifont}
\usepackage{fp}
\usepackage{subcaption,multirow,xspace,makecell}
\usepackage{amsfonts}
\usepackage{soul}
\usepackage[shortlabels]{enumitem}

\newcommand{\newfpeval}[1]{\FPeval{\result}{#1}\result}

\newcommand*\mininumbercircled[1]{\tikz[baseline=(char.base)]{
		\node[fill=darkgray, text=white, shape=circle,draw,inner sep=0.1pt] (char) {#1};}}

\newcommand{\GNNTrainer}{{GNNDrive}\xspace}

\AtBeginDocument{
  }

\copyrightyear{2024} 
\acmYear{2024} 
\setcopyright{rightsretained} 
\acmConference[ICPP '24]{The 53rd International Conference on Parallel Processing}{August 12--15, 2024}{Gotland, Sweden}
\acmBooktitle{The 53rd International Conference on Parallel Processing (ICPP '24), August 12--15, 2024, Gotland, Sweden}
\acmDOI{10.1145/3673038.3673063}
\acmISBN{979-8-4007-1793-2/24/08}

\begin{document}

\title[Reducing Memory Contention and I/O Congestion for Disk-based GNN Training]{Reducing Memory Contention and I/O Congestion\\ for Disk-based GNN Training}

\author{Qisheng Jiang}
\affiliation{
	\institution{ShanghaiTech University}
	\city{Shanghai}
	\country{China}
}
\orcid{0000-0002-5570-0018}

\author{Lei Jia}
\affiliation{
	\institution{ShanghaiTech University}
	\city{Shanghai}
	\country{China}
}
\orcid{0009-0004-1601-2695}

\author{Chundong Wang}
\authornote{
This is a full version for the paper with almost the same title accepted by the  53rd International Conference on Parallel Processing (ICPP '24).
	C. Wang is the corresponding author (cd\_wang@outlook.com). 
The source code is available at \url{https://github.com/toast-lab/GNNDrive}.}
\affiliation{
	\institution{ShanghaiTech University}
	\city{Shanghai}
	\country{China}
}
\orcid{0000-0001-9069-2650}

\begin{abstract}
	Graph neural networks  (GNNs) gain 
	wide popularity. 
	Large graphs with high-dimensional
	features 
	become 
	common 
	and training GNNs on them is 
	non-trivial on an ordinary  machine. 
	Given a gigantic graph, 
	even 
	sample-based GNN training cannot 
	work efficiently, since it is difficult to keep the graph's entire data  in  memory
	during
	the training process. 
	Leveraging a solid-state drive (SSD) or other storage devices to extend the memory space
	has been studied 	in training GNNs.
	Memory and I/Os are hence critical for effectual disk-based    training.
	We find 
	that state-of-the-art (SoTA) disk-based GNN training systems
	severely suffer from issues like the 
	 memory contention between a graph's topological  and feature data,
	and severe I/O congestion upon loading data
	from SSD for training. 
	We accordingly 
	develop \GNNTrainer.
	\GNNTrainer 1)
	minimizes the memory footprint with holistic buffer management
	across sampling and extracting,
	and 2) avoids I/O congestion 
	through
	a strategy of asynchronous feature extraction. 
	It also avoids costly data preparation on the critical path
	and makes the most of software and hardware resources.
	Experiments show that \GNNTrainer
	achieves superior performance.
	For example, when training with the
	Papers100M dataset and
	GraphSAGE model, 
	\GNNTrainer is faster than SoTA PyG+, Ginex, and MariusGNN by 
	16.9$\times$, 2.6$\times$,
	and 2.7$\times$, respectively.

\end{abstract}



\maketitle

\section{Introduction}\label{sec:intro}

Graph Neural Networks (GNNs)
demonstrate remarkable outcomes in  workloads 
such as recommendation~\cite{gnn:Survey:Recommender,gnn:Web-Scale-Recommender}, traffic prediction~\cite{gnn:traffic}, fraud detection~\cite{gnn:Fraud,gnn:GraphScope}, and drug discovery~\cite{gnn:drug}.
By leveraging both node-level features and structural graph information,
 GNNs 
capture local topology and generate embeddings 
that are valuable for downstream tasks~\citep{GNN:multi-core:ICPP-2021,gnn:gnnlab,ICPP:MG-GCN,gnn:PaGraph,9252000,9680339,gnn:hongtu:sigmod23}.

While GNN models are effective in practice, 
training them poses both implementation and performance challenges 
within existing tensor-oriented frameworks~\cite{tool:pytorch,tool:tensorflow}. 
These challenges stem from the recursive feature update mechanism, 
which requires aggregating information from neighboring nodes in an input graph~\cite{gnn:gnnlab}. 
Covering all neighbors in 
a certain range 
for each training node becomes 
infeasible, due to
the high computational cost and memory requirements~\cite{sample:scale}. 
Sample-based GNN training, which selects a 
number of 
neighbors  by sampling in 
a specific range (hops) for a training node, 
has emerged as a practical and viable solution~\cite{gnn:gnnlab,gnn:bytegnn}. 
In short, it divides all training nodes into {\em mini-batches}, each of which is a subset of the entire training dataset, 
and 
iteratively trains a GNN model in a  {\bf s}ample, {\bf e}xtract, and {\bf t}rain (SET)  fashion 
during one {\em epoch}, i.e., a complete cycle of training the whole  dataset. 
The strategy of sampling 
enables 
significant reduction in computations~\cite{gnn:gnnlab,gnn:ginex}.

In GNN training systems, 
the entire dataset of a graph  is  
supposed to be buffered in  memory throughout   the training process. However,
large graphs with high-dimensional features turn to be  widespread. 
Both topological data and node features   are rapidly expanding~\cite{dataset:ogblsc,gnn:bytegnn}. 
Regarding the physical memory capacity,
even sample-based  training   encounters
practical difficulty of dealing with massive graph data 
on a  machine.
How to efficiently train GNNs with large graphs is
a considerable challenge today.

Distributed GNN training with a cluster of machines exploits collective computation powers and memory capacities to handle large graphs~\cite{gnn:bytegnn,gnn:AliGraph}.
However, it 
may 
entail underutilized hardware and high 
monetary cost. 
Disk-based   training is the other  promising method with  low  cost~\cite{gnn:ginex,gnn:mariusgnn}. 
Using the capacity and performance potentials offered by storage devices such as a solid state drive (SSD), 
disk-based GNN training 
can overcome the limitation of memory capacity 
and enable efficient processing of gigantic 
datasets   on one 
machine. 
Though, effectual disk-based GNN training is non-trivial, especially
on an ordinary machine with moderate hardware components such as memory, GPU, and SSD
that small and midsize
enterprises as well as academic researchers are usually using.

We have conducted a study with three state-of-the-art  (SoTA)
disk-based GNN training systems, i.e.,   PyG+~\cite{gnn:ginex}, 
Ginex~\cite{gnn:ginex}, and MariusGNN~\cite{gnn:mariusgnn},
on an ordinary machine with prevalent GNN models and graph datasets. 
Training results 
show that
they severely suffer from issues caused by memory and I/Os.
PyG+ exhibits serious memory contention between 
topological and feature data used for sampling and extracting, respectively.
When PyG+ does  both  sampling and  extracting instead of sampling only,
the time spent on sampling can rise by 5.4$\times$.
It also experiences substantial I/O congestion due to synchronously loading data from SSD between training mini-batches.
Ginex alleviates memory contention with separate caches dedicated for sample and extract stages,  
but still shares with PyG+ the issue of  I/O congestion upon loading data for training. 
MariusGNN splits a graph into partitions and 
minimizes I/Os by training on partitions that have been buffered in memory.
It thus  reduces I/O congestion in a training epoch. 
However, MariusGNN introduces long I/O wait time for data preparation, in which
it makes a sequential order of multiple partitions and loads corresponding partitions into memory in advance
for a proper training outcome.
Such data preparation occurs on the critical path of training for each epoch and can take up to
46.1\

Motivated by these observations, we develop \GNNTrainer, a disk-based   training system  
specifically 
designed to optimize training with large graphs 
and ordinary hardware. 
\GNNTrainer minimizes the memory footprint used for feature extraction, in order to reduce memory contention between sampling and extracting. 
It employs asynchronous feature extraction to mitigate I/O congestion. 
It also avoids costly
 data preparation prior to training.
Additionally,  \GNNTrainer makes use of concurrent threads and multiple GPUs if available,
thereby further enabling out-of-order execution and data parallelism, respectively,
for performant GNN training.

To sum up, we make following contributions in this paper. 
\begin{itemize}[leftmargin=4mm]\setlength{\itemsep}{-\itemsep}
    \item We look into disk-based GNN training with a study on SoTA systems.
    We  find that
    the memory contention and I/O congestion
    are main factors that impair
    the efficiency of them.
    We accordingly propose a novel GNN training system named \GNNTrainer.
	\item \GNNTrainer reduces the memory contention across stages.
	It minimizes the memory footprint of buffers employed to extract feature data and efficiently accommodates topological data in host memory for sampling. It also avoids consuming the page cache of operating system (OS) through direct I/Os.
 	  \item As to I/Os, for a graph's feature data,  \GNNTrainer schedules the   loading from SSD to host memory and the transfer from host memory to device memory to be both asynchronous. It overlaps feature extractions for one mini-batch
 	  with training other mini-batches  to be parallel at runtime, thereby hiding I/O wait time.
    \item \GNNTrainer avoids costly data preparation. It 
     considers mini-batch reordering to execute training out of order. It can
    also  parallelize a training task by exploiting multi-GPUs. 
\end{itemize}

We prototype 
\GNNTrainer on PyTorch Geometric (PyG)~\cite{gnn:pyg}.
We
evaluate it with three prevalent models 
over four real-life  graphs.  
Extensive experiments 
show that \GNNTrainer yields
substantially high performance.
For example,    when training with 
Papers100M dataset~\cite{dataset:ogb}  and 
GraphSAGE model~\cite{model:sage}, 
\GNNTrainer is 16.9$\times$, 2.6$\times$, and 2.7$\times$ faster than 
PyG+, Ginex, and MariusGNN, respectively.

\section{Background}\label{sec:background}

{\bf GNN.}
GNN operates on a graph $G=(V, E)$, where each node $v$ is associated with a feature vector. 
The objective of GNN  is,  for a target node, to generate
embeddings that capture both the node's individual features and the information from its $k$-hop in-neighbor nodes. 
A GNN model comprises multiple layers. 
Each layer executes two primary steps: aggregation and combination. 
In the aggregation step, the features of the incoming nodes are joined into a single vector using aggregation functions like mean, max,  sum,
or more advanced  functions~\cite{model:gat}.  
Then the aggregated feature undergoes the combination step that
 applies a non-linear function through a fully connected (FC) layer. 
These two steps are
repeated for $k$ layers 
 to capture information from up to $k$-hop in-neighbors.

\begin{figure}[t]
	\centering
	\includegraphics[width=\linewidth,page=1]{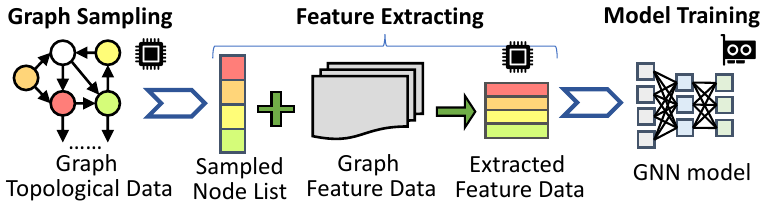}
	\caption{An example of sample-based GNN training system.}\label{fig:gnn}
\end{figure}

{\bf Sample-based Training.}
It is increasingly common to encounter large graphs with high-dimensional features~\cite{gnn:bytegnn}.
For instance, MAG240M, 
which is part of the widely-used Open Graph Benchmark (OGB)~\cite{dataset:ogb,dataset:ogblsc},
consists of 240 million nodes with 768-dimensional features. 
Training GNNs at such a large scale 
presents challenges with regard
 to the high computational cost and memory requirements~\cite{sample:scale}. 
To address these issues, sample-based training has 
 emerged~\cite{model:sage,gnn:mariusgnn}. 
As depicted in~\autoref{fig:gnn}, it involves dividing the training nodes into
 mini-batches and iteratively proceeds 
 in  {\bf s}ample, {\bf e}xtract, and {\bf t}rain (SET)  steps~\cite{gnn:gnnlab}. 
In the sample stage, an input graph is sampled based on a user-defined algorithm 
that takes into account the graph's topological data and
produces a list of sampled nodes.
Next, the features of sampled nodes are extracted into a separate buffer. 
Finally,  training is performed 
on extracted features in the train stage.
For GPU-based GNN training, 
the extracted features must be transferred to the GPU's device memory before training in the train stage.
Prior studies have demonstrated that sample-based training substantially reduces computational costs and achieves comparable accuracy to full-batch training~\cite{gnn:gnnlab}. 
Therefore, it
gains increasing popularity,
especially to handle large graphs~\cite{gnn:gnnlab,gnn:bytegnn,gnn:pyg}.

{\bf Disk-based GNN Training.}
The topological and feature data of large graphs are likely to exceed the 
memory capacity of a  machine,
especially for ordinary, economical ones.
For example, 
the paper nodes with float32 feature data
in MAG240M 
demand about 
350GB of memory space. 
Graphs found in production environments 
are even larger~\cite{gnn:bytegnn,gnn:ginex}. 
A solution to tackle a large graph
is to partition 
and distribute subgraphs 
across multiple machines for 
 training~\cite{gnn:AliGraph,gnn:bytegnn}. 
However, 
the cost of purchasing and managing a cluster of machines is   concrete, while the efficiency may not be high~\cite{gnn:mariusgnn,gnn:ginex}. 
Disk-based GNN training stands out as a promising solution 
by storing all data on disk and loading the data to be used into memory on demand for training on 
a target subgraph~\cite{gnn:mariusgnn,gnn:ginex}.
By doing so, it exploits
 the ample capacity and cost efficiency of disk drives~\cite{gnn:ginex}.

\begin{figure*}[t]
	\begin{minipage}[t]{0.31\textwidth}
		\includegraphics[width=\linewidth,page=1]{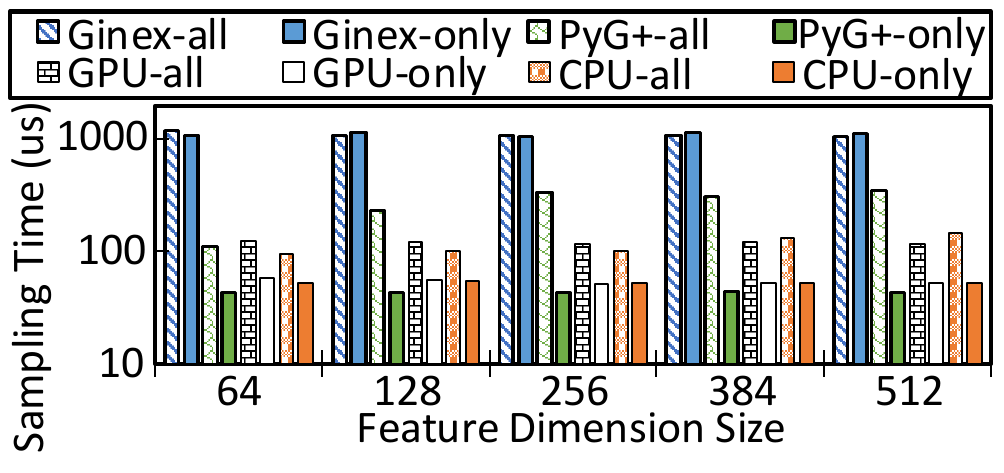}
		\caption{Sampling time for Ginex, PyG+, and \GNNTrainer in  varying feature dimension sizes.}
		\label{fig:mot-sample:dim}
	\end{minipage}
	\hspace{0.004\textwidth}
	\begin{minipage}[t]{0.67\textwidth}
		\centering
		\vspace{-19ex}         
		\begin{subfigure}{0.32\linewidth}
			\includegraphics[width=\linewidth,page=1]{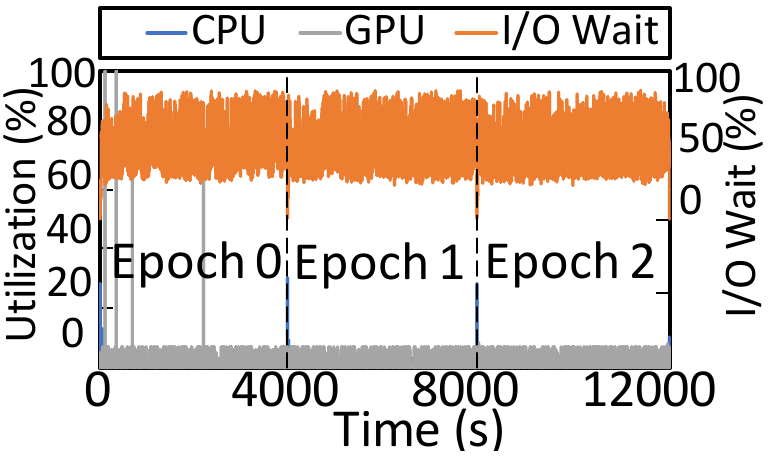}
			\caption{PyG+.}
			\label{fig:mot-io:pyg}
		\end{subfigure}	
		\begin{subfigure}{0.33\linewidth}
			\includegraphics[width=\linewidth,page=2]{mot-io.pdf}
			\caption{Ginex.}
			\label{fig:mot-io:ginex}
		\end{subfigure}
		\begin{subfigure}{0.33\linewidth}
			\includegraphics[width=\linewidth,page=1]{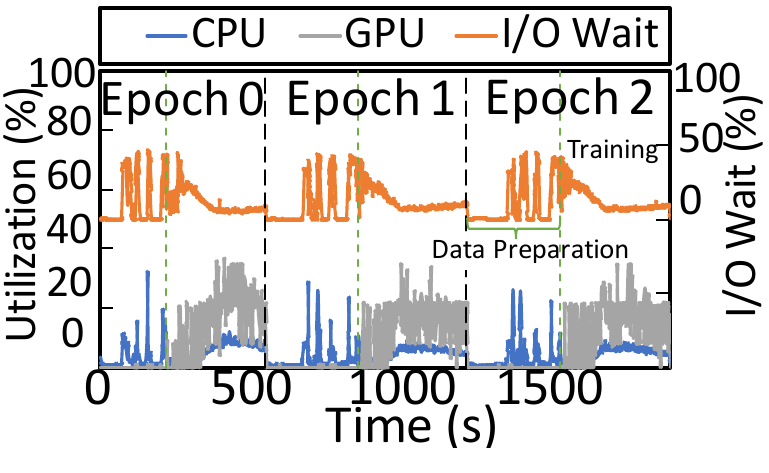}
			\caption{MariusGNN.}
			\label{fig:mot-io:marius-32}
		\end{subfigure}	
		\caption{CPU utilization, GPU utilization and ratio of I/O wait time with PyG+, Ginex, and MariusGNN for a time window of three epochs.}\label{fig:mot-io}
	\end{minipage}
\end{figure*}

One straightforward way to make disk-based GNN training is to extend the original training system 
by directly using memory-mapped graph data. For example, 
PyG+ 
memory-maps
topological and feature data for a graph and converts them into PyTorch tensors~\cite{gnn:pyg,gnn:ginex}.
Another way is to 
customize in-memory caches to buffer and manage graph data,
instead of relying on the page cache of  OS. 
Ginex~\cite{gnn:ginex} follows this way.
It has one neighbor cache for topological data and the other feature cache. 
It further restructures the   training procedure  
by separating  sample and extract stages,  thereby
allowing for
an optimized replacement algorithm 
to cache feature vectors 
and in turn reduce I/Os. Moreover,
Ginex analyzes  sampling results in an initial phase 
to prepare 
the caching 
for subsequent extractions. 
MariusGNN~\cite{gnn:mariusgnn} takes a third way that
partitions a graph
and transfers subsets of these partitions to host memory for training.
It avoids swapping partitions in an epoch to minimize I/Os. 
Note that MariusGNN orders partitions in a desired sequence 
and replaces them between epochs to maintain training accuracy.
This   causes compulsory data preparation before training.
Worse, 
MariusGNN assumes a risk of undermining the accuracy of training models~\cite{gnn:mariusgnn}, 
as it samples nodes solely with buffered  partitions.

\section{Motivation}\label{sec:motivation}

GNN training time dramatically increases with the scale-up of
graphs. 
PyG+, Ginex, and MariusGNN are SoTA disk-based GNN training systems.
We have conducted a comprehensive study
 to analytically figure out if they 
suffer from performance overheads that lead to
suboptimal training efficacy.
We aim to 
 find out
the radical challenges and limitations that affect disk-based   training.
We utilize a 3-layer GraphSAGE model with a sample size of (10, 10, 10) for the Papers100M dataset
with ordinary hardware resources (e.g., memory, SSD, and GPU). 
We configure the default capacity of host 
memory as 
 32GB, 
with regard to the sizes and scales of datasets that are publicly available. By default,
the node feature has a  dimension size of 128.
Details of testing setup and methodology are shown in Section~\ref{sec:eval}.

{\em The focus on sample and extract stages}.
We firstly analyze the breakdown of disk-based GNN training with sampling to identify the bottleneck among 
SET stages.
On training with
Papers100M, 
the extract stage accounts for 97.3\
while the sample stage  takes 1.0\
As a result,
these two stages constitute the majority of total training time. 
Our   analysis hence concentrates on them.
In brief,
we find that PyG+, Ginex, and MariusGNN severely suffer from  performance overheads at different aspects.

$\mathfrak{O}1$ {\em Memory Contention.} 
\textbf{Memory contention incurs severe inefficiency to disk-based GNN training}, particularly for sampling. 
We quantitatively measure
the sampling time 
with varying feature dimension sizes. 
We consider two scenarios: 
1) when only the sample stage is performed per training epoch for each system, denoted with a suffix `-only'   (e.g., PyG+-only),
and 2) when 
all SET stages are performed per epoch, denoted
with a suffix `-all' (e.g., PyG+-all). 
The results are shown in~\autoref{fig:mot-sample:dim}.
Note that the 
Y-axis is 
in a logarithmic scale. 
We can obtain following observations.

\textit{Firstly, memory contention arises due to the interaction between topological and feature data}, especially for PyG+.
The sampling time of PyG+-only is significantly smaller than that of PyG+-all. 
Given the default dimension size of 128,
PyG+-all results in  5.4$\times$ sampling time 
compared to PyG+-only at the sample stage. 
This discrepancy is due to memory contention caused by
the feature data used in the extract stage against the topological data required for the sample stage.
For PyG+,
both types of data are memory-mapped into host memory and compete the OS's page cache (see Section~\ref{sec:background}).
Worse, the concurrent execution of sample and extract stages in PyG+ 
exacerbates the problem of memory contention. 

As to Ginex,
the sampling time of Ginex-only is close 
to that of Ginex-all. 
 Ginex 
utilizes separate neighbor and feature caches for sample and extract stages, respectively. 
Furthermore, Ginex conducts sampling  in advance using a superbatch, 
which is a bundle of many mini-batches (1,500 by default).
This
further helps to relieve the memory contention between   sample and extract stages.

\textit{Secondly, datasets with higher feature dimensions cause even worse 
	 memory contention.}
Higher dimensions mean
more memory space to be consumed in the extract stage and interfere  more with   sampling, 
particularly for memory-mapped-based PyG+.  
When handling feature data with, say,  a dimension size of 512, PyG+-all
spends 3.1$\times$ sampling time compared to that with a dimension size of 64.

$\mathfrak{O}2$ {\em I/O congestion.}
\textbf{I/O congestion occurs when SoTA systems are interacting with SSD.}
	{Tedious I/O wait  even stalls CPU or GPU from computing}.
We have monitored 
the ratios of
CPU and GPU utilizations versus the ratio of I/O wait time 
in three consecutive 
epochs for PyG+, Ginex, and MariusGNN. 
As shown in Figures 3\subref{fig:mot-io:pyg} and 3\subref{fig:mot-io:ginex},
when the ratio of I/O wait time is high for  PyG+ and Ginex, the utilization ratios of CPU and GPU are very low.
This is due to the long synchronous loading of data from  SSD.
Both of them experience substantial I/O wait time. Meanwhile, CPU and GPU have to stay idle for the readiness of data.
PyG+ heavily relies on memory-mapped I/Os between SSD and the OS's page cache. 
 Ginex's feature cache and replacement policy effectively mitigate excessive I/Os in the train stage. 
Though, it still undergoes I/O congestion for sampling 
as well as the initialization of feature cache at the start of each superbatch.
Moreover, 
 it  stores sampling results into  SSD 
per
 superbatch 
to optimize the replacements for its feature cache. This
costs
extra I/Os  and longer sampling time.

Unlike PyG+ and  Ginex, MariusGNN splits a graph into  partitions. 
At runtime,
it mainly utilizes data of a buffered partition for sampling and extracting, thereby minimizing I/O overheads in an epoch.
As a result,
MariusGNN indeed reduces I/O wait time during training (see Figure 3\subref{fig:mot-io:marius-32}). 
However, it   suffers from intense I/O wait time for data preparation.
To guarantee correct training outcome and high efficiency, 
MariusGNN prepares data by  ordering a desired sequence of  partitions 
 and loading them to memory in advance
for an epoch. 
With 32GB host memory, data preparation takes as much as 46.1\
due to 
 I/O congestion. 
This I/O overhead is dramatic but distinct from those 
with PyG+ and Ginex.

\textbf{Design Decisions}. 
With $\mathfrak{O}1$ and $\mathfrak{O}2$, 
SoTA disk-based GNN training systems face 
different and serious performance issues in handling large graphs, especially on a   machine with ordinary hardware. 
 Such  machines  are common
and affordable for academic researchers as well as small and medium enterprises. 
We intend to develop a new GNN training system to overcome aforementioned challenges and make the most of 
an ordinary machine,
without losing correctness or accuracy. 
Our design decisions are summarized as follows.
\begin{itemize}[leftmargin=4mm]\setlength{\itemsep}{-\itemsep}
	\item We need a holistic inter-stage strategy to minimize the memory contention between topological   and feature data to be used in sample and extract stages, respectively.
	\item We shall reshape the I/O model to reduce the cost of synchronously loading and transferring data in the training process.
	\item We should avoid heavy data preparation 
				when systematically scheduling data and operations  to make use of software (e.g., multi-threads) and hardware (e.g., multi-GPUs)
				for parallelism. 
\end{itemize}

\section{Design of \GNNTrainer}\label{sec:design}

\GNNTrainer is an 
efficient and flexible system, specifically 
designed to optimize the sample-based training of GNNs on large-scale graphs 
through a coordinating utilization of CPU, GPU, memory, and SSD on a  machine.
\GNNTrainer 
 focuses 
on 
two aforementioned 
challenges,
i.e., memory contention and I/O congestion. 
To address them, 
\GNNTrainer 
1) 
more supports sampling through a systematic inter-stage buffer management scheme, 
in order to minimize the memory footprint for training,
and 2) conducts asynchronous feature extraction by scheduling I/O and memory operations
outside of the critical path. 
Additionally, 
it exploits software (threads) and hardware (GPUs) 
to gain higher  efficiency.

\subsection{Overview}

\GNNTrainer stores
both topological and feature data of a graph  in SSD.
It organizes
each node's feature data in ascending order of node IDs 
to make 
a  table. 
It forms the topological data of a graph as an
adjacency matrix
for efficient   sampling. 
\autoref{fig:arch-gpu} illustrates 
how \GNNTrainer samples and trains with CPU and GPU, respectively.
Four main stages form its entire process, i.e., 
sampling, extraction, training, and releasing.
In sampling a graph,
\GNNTrainer employs a pool of   threads as {\em samplers}
(\mininumbercircled{1} in~\autoref{fig:arch-gpu}). 
Each sampler generates a subgraph
 for a mini-batch from which 
a 
list of  sampled node IDs is made.  This list of nodes is
to be enqueued into the extracting queue 
(\mininumbercircled{2}). 
In the subsequent feature extraction stage, 
every extracting thread ({\em extractor}) dequeues a sampled node list from the extracting queue 
(\mininumbercircled{3})
to asynchronously load and transfer
the corresponding feature data from SSD into the device memory of GPU
(\mininumbercircled{4} and \mininumbercircled{5}).

Once an extraction completes,
the extractor enqueues a node alias list 
into the training queue
(\mininumbercircled{6}). 
To
train the model,  
the training stage 
utilizes the sampled subgraphs dequeued from the training queue 
and corresponding feature data already loaded into GPU's device memory 
(\mininumbercircled{7}). 
The training thread ({\em trainer}) adopts the node alias list to index required feature data in the device memory
and   starts 
training 
with indexed feature data.
After training, the original sampled node list, as already processed in extract and train stages, 
is enqueued into the releasing queue
(\mininumbercircled{8}). 
Finally,
in the releasing stage,
the releaser dequeues sample nodes from the releasing queue and
frees up
memory space   for the next iteration
(\mininumbercircled{9}).

Three queues 
play  the middle-person roles 
 between stages.
As simultaneously used over time, they 
enable a pipelined effect over four stages. 
Also, they do  not pose
any bottleneck, because everyone of them only deals with sampled node IDs, rather than 
 actual data.

\subsection{Asynchronous Feature Extraction}\label{sec:async-extract}

\begin{figure}[t]
	\centering
	\scalebox{0.92}{\includegraphics[width=\linewidth,page=1]{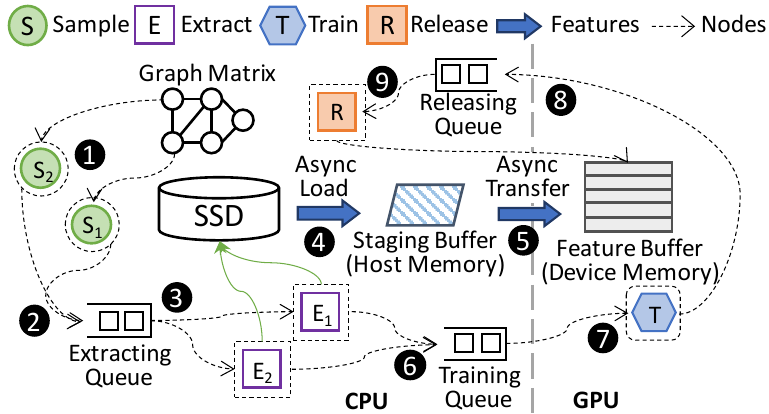}}
	\caption{An illustration of \GNNTrainer's architecture.}\label{fig:arch-gpu}
\end{figure}

\textbf{Reduced Memory Footprint.}
As the inter-stage memory contention more impacts graph sampling than feature extraction,
\GNNTrainer particularly minimizes 
the memory footprint used for the latter.
As shown at the right of \autoref{fig:arch-gpu},
\GNNTrainer allocates a {\em feature buffer} in GPU's device memory to store the feature data for training. 
In the center of \autoref{fig:arch-gpu},
\GNNTrainer manages a {\em staging buffer} in the host memory, solely used for 
transferring feature data from SSD to the feature buffer. 
The size 
of staging buffer is 
bounded
by the number of extractors 
and the number of features 
to be loaded to GPU for each extractor. 
Therefore, the staging buffer can be expanded or shrunk  by adjusting the number of extractors,
which we decide with regard to the volume of topological data and the capacity of available host memory.
The size of feature buffer is determined by the depth of training queue
and the number of nodes involved in a training mini-batch. 
As the extracted nodes in the training queue consume the GPU's device memory,
this queue's depth
 is restricted by the capacity of device memory to avoid the out-of-memory (OOM) issue during training.
As justified by our
 evaluation in Section~\ref{sec:eval:deepdive}, 
both feature and staging buffers consume less memory space 
compared to those buffers  with similar purposes used by SoTA systems, e.g., Ginex.
Moreover, \GNNTrainer employs direct I/Os that bypass the OS's page cache 
to load feature data from SSD, bypassing the OS's page cache. 
The reason for doing so is twofold. Firstly,
by eliminating the need for OS's pages,
\GNNTrainer
further reduces the memory footprint for extracting.
Secondly, on loading data with a large I/O depth,
direct I/Os yield comparable performance than buffered I/Os (see~\autoref{app:async-compare}).

\textbf{Feature Buffer Management.} 
\GNNTrainer has 
a lightweight mechanism 
to manage the feature buffer. 
It comprises four  components: 
a mapping table, a buffer, a reverse mapping array, and a standby list (see \autoref{fig:arch-extract}). 
An entry in the
mapping table keeps the mapping metadata from 
a graph node to a buffer slot, 
including 
 slot index, reference count, and valid bit for a node. 
The slot index refers to the mapped slot or $-1$ (`{\em not applicable}').
 The reference count tracks if 
 the data can be released from the feature buffer or not. 
The valid bit and slot index 
jointly
tell if data has been   extracted into a   buffer slot or not. 
If the slot index is not 
$-1$ and valid bit is `1', 
the node's   data is ready in the slot.
If the slot index is not 
$-1$ and valid bit is `0',
the data is being extracted into the slot.
With $-1$ slot index  and `0' valid bit,
the  data is not in the feature buffer.
The case of 
`not applicable' slot index  and `1' valid bit
is impossible. 

\GNNTrainer has a {\em standby} list to group free slots
as well as ones that have retired for past mini-batches but are likely to be reused.
It uses 
a hash table 
to track and manage 
these standby slots 
 in the least-recently-used (LRU) way. 
\GNNTrainer employs
the reverse mapping array to track the mapping from slot indexes to node IDs
in order to quickly identify the node being buffered in a specific slot.
When no valid node is stored in a slot, it puts $-1$ in the entry.

\begin{figure}[t]
    \centering
    \includegraphics[width=\linewidth,page=1]{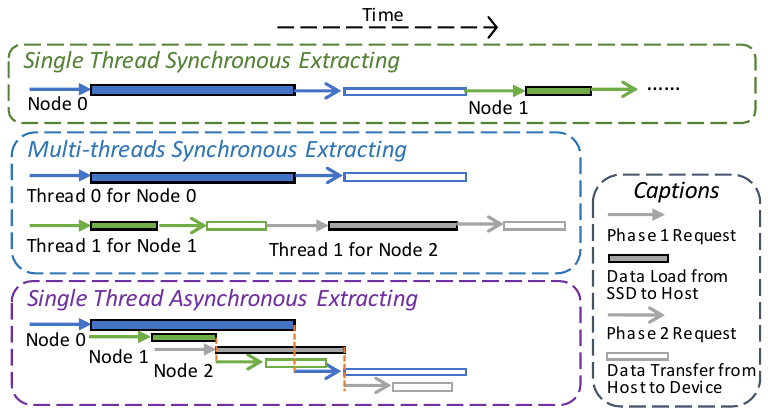}
    \caption{A comparison between asynchronous and synchronous extracting.}\label{fig:arch-async}
\end{figure}

\textbf{Asynchronous Extracting.} 
\GNNTrainer maintains a pool of extractors.
It utilizes them  to concurrently extract feature data for different training mini-batches. 
Each extractor 
performs a two-phase extraction
with SSD  for a mini-batch
and handles both phases in an asynchronous fashion.
An I/O request loads the required feature data for one node at a time. 
In the first phase, the extractor issues  I/O requests to load data 
into the staging buffer 
for all the nodes involved in one 
mini-batch (\mininumbercircled{4} in \autoref{fig:arch-gpu}). 
The extractor 
constructs and issues multiple asynchronous I/O requests that can be enabled by the latest asynchronous I/O libraries, such as the io\_uring~\cite{tool:iouring} (see \autoref{app:async-intro} for more detail).
It later checks and collects the return value per request, without synchronously waiting on the critical path.
Once the loading 
for a node succeeds, 
the extractor proceeds to the second phase, 
where it transfers the node's feature data from the staging buffer to  feature buffer (\mininumbercircled{5} in \autoref{fig:arch-gpu}). 
It does not wait to initiate a transfer on the completion of loading all nodes,
but 
launches a transfer once a node is  loaded.
\GNNTrainer leverages the CUDA Toolkit~\cite{tool:cuda} to transfer data asynchronously from host memory to GPU's device memory.
Therefore,
besides parallelizing I/Os,
\GNNTrainer further 
parallelizes the loading of   current node
with the transfer of previous node in each mini-batch.

We have done a test by
comparing synchronously loading massive data with multiple threads
against asynchronously loading data with one  thread.
They achieve a similar bandwidth (see~\autoref{app:async-compare}). Hence, 
as shown by 
~\autoref{fig:arch-async}, 
\GNNTrainer dedicates 
one extractor to a mini-batch,
 handling the feature extraction asynchronously. 
This also eliminates intra-thread context switches in extracting feature data within one mini-batch. 
In summary, the asynchronous extraction highly reduces wait time 
by efficiently reshaping and rescheduling the flows of I/O and memory operations. 
With it,
\GNNTrainer
manages to overcome I/O congestion and enables
CPU cores 
to be utilized for other tasks such as sampling or training.

 \textbf{Extraction Procedure.}
At the beginning of 
extraction, 
the extractor dequeues the sampled node list for a 
mini-batch 
from the extracting queue (see \mininumbercircled{3} in
\autoref{fig:arch-gpu}). 
\GNNTrainer 
manages a node alias list that holds slot indexes in the feature buffer
for aliasing nodes.
It sets the initial aliases for all nodes to be $-1$s.
The extractor also has a wait list for nodes 
that other extractors are extracting.

Next, the extractor checks if a node's data is already present in the feature buffer 
by examining the node's valid bit in its mapping table entry. 
By doing so, \GNNTrainer aims to reuse data that has been extracted.
If the valid bit is `1', 
the extractor checks the node's reference count. 
A zero reference count implies that
the relevant 
buffer slot is in the standby list. 
The extractor fetches the slot and places 
the slot's index 
in the node alias list for that node. 
Note that
the reference count of an invalid node (with valid bit being `0') 
might be greater than 0. 
This occurs when the  feature data is still being extracted by another thread. 
The extractor inserts such nodes into the wait list to avoid redundant extraction. 
It re-examines the valid bits of those nodes at the end of current extraction
to confirm the data's readiness for use.
Additionally,
for sampled nodes, the extractor increments their reference counts.

Then, 
\GNNTrainer asynchronously loads   required data from SSD
into the feature buffer (\mininumbercircled{4} in \autoref{fig:arch-gpu}). 
The extractor efficiently skips data that is already present in the feature buffer by checking the node aliases. 
Before loading a node, 
the extractor selects the LRU slot in the standby list. 
If the reverse mapping of the LRU slot is greater than $-1$, 
the slot is not empty and contains valid data of previous node. 
The extractor invalidates that previous node's mapping entry 
by resetting the node's valid bit to `0'. 
If  no slot is available in the standby list, 
the extractor waits for the completion of releasing used nodes. 
To avoid deadlock, a minimum of $N_e \times M_b$ slots are reserved 
in the feature buffer,
where $N_e$ represents the number of extractors 
and $M_b$ is the maximum number of nodes in a training mini-batch. 
After slot allocation,
the extractor updates the
mapping table, reverse mapping array, and node alias list.
Then the extractor submits an asynchronous loading task to the staging buffer for the node 
and proceeds to handle the next node without waiting for the completion of submitted   task.

When all loading I/O tasks are submitted, 
the extractor waits for the completion signals. 
Once a node is loaded,
the extractor submits an asynchronous transfer task 
from   staging buffer to   feature buffer (\mininumbercircled{5} in \autoref{fig:arch-gpu}). 
Likewise, the extractor does not wait for the completion of transfer 
but continues to transfer the next loaded node. 
When all 
transfer tasks are  submitted, 
the extractor waits for the completion signals. 
After transferring a node, 
the extractor sets the node's valid bit to `1' 
to notify other threads that the node's data has been  extracted into the feature buffer. 
As mentioned,
an extractor hence  shares and reuses nodes with other extractors, without incurring redundant I/Os.
Through node aliasing, 
the extractor eventually obtains the node alias list of a mini-batch 
for training.
We stepwise present the  extraction procedure  with Algorithm \autoref{algo:extract} in \autoref{app:algorithm}.

\textbf{Release Feature Buffer.} As shown by \mininumbercircled{8} in 
\autoref{fig:arch-gpu}, \GNNTrainer enqueues
the original sampled node list 
in the releasing queue.
The releaser dequeues such a list and
for each involved node, 
it decrements the node's reference count. 
If the reference count becomes zero, 
the releaser adds the corresponding slot to the standby list. 
The invalidation of  original node's mapping table entry is delayed 
until the slot is to be used by other nodes. This
promotes the potential reuse of a node with regard to inter-batch locality.

\begin{figure}[t]
	\centering
	\includegraphics[width=0.95\linewidth,page=1]{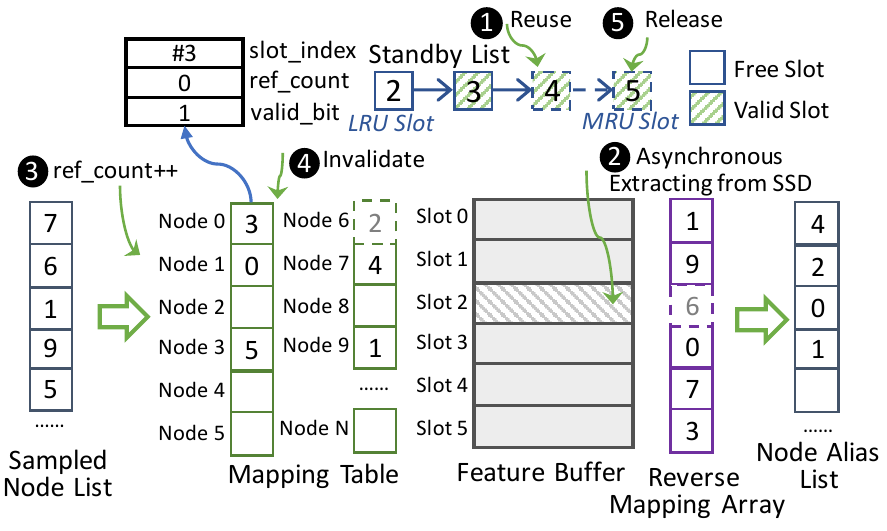}
	\caption{An example of \GNNTrainer's extract stage.}\label{fig:arch-extract}
\end{figure}

\textbf{End-to-end Process.} 
\autoref{fig:arch-extract} exemplifies the asynchronous  extraction of \GNNTrainer. 
The first element in the sampled node list is node 7. 
The extractor checks the mapping table and finds that the data for node 7 is valid in slot 4 with a zero reference count. 
To reuse slot 4, the extractor obtains it from the standby list 
(\mininumbercircled{1} in~\autoref{fig:arch-extract}). 
The extractor hence sets the node alias of node 7 to 4. 
Since node 6 is not valid in the mapping table and slot 2 is the first available free slot,
the extractor asynchronously extracts the data for node 6 from SSD to slot 2 
(\mininumbercircled{2}). 
The reverse mapping of slot 2 is set to 6 after extraction. 
Both nodes 1 and 9 are valid. As their reference counts are greater than zero, 
the data for them must have been enqueued into the training queue or
are being used by other threads for extraction or training. 
The extractor simply increases their reference counts 
(\mininumbercircled{3}). 
Similar to node 6, node 5 is also not valid in the mapping table. 
The extractor allocates for it the next LRU slot, i.e., slot 3, which is still valid but not being used. 
To invalidate slot 3's original node 0,
it clears the node's valid bit to be `0'
(\mininumbercircled{4}). 
The extractor then performs asynchronous extraction for node 5. 
After obtaining the node alias list of  mini-batch, 
the extractor enqueues the node aliases into the training queue for training. 
In the release stage, 
the releaser decrements the reference counts for nodes. As the reference count of 
node 3 becomes zero,
at the tail of standby list, 
the releaser adds slot 5 to which node 3 is mapped 
(\mininumbercircled{5}).

\subsection{Reordering and Parallelism}

\textbf{Mini-batch Reordering.}
\GNNTrainer decouples sampling from extracting. It employs multiple threads 
in both stages for parallel processing.
Samplers concurrently deals with multiple mini-batches. 
They may finish  at different paces due to variations in between. 
Thus, the order of enqueuing mini-batches into the extracting queue 
may differ from the initial order of mini-batches. 
Similarly, 
extractors may 
enqueue extracted mini-batches into the training queue out of order.
This  {\em   reordering} helps to prevent performance  degradation 
caused by occasional long latencies 
for sampling or tedious 
extraction on some irregular mini-batches. 
We also justify that reordering
does not affect convergence or   accuracy (see Section~\ref{sec:eval:convergence}).

\begin{figure}[t]
	\centering
	\scalebox{0.99}{\includegraphics[width=\linewidth,page=1]{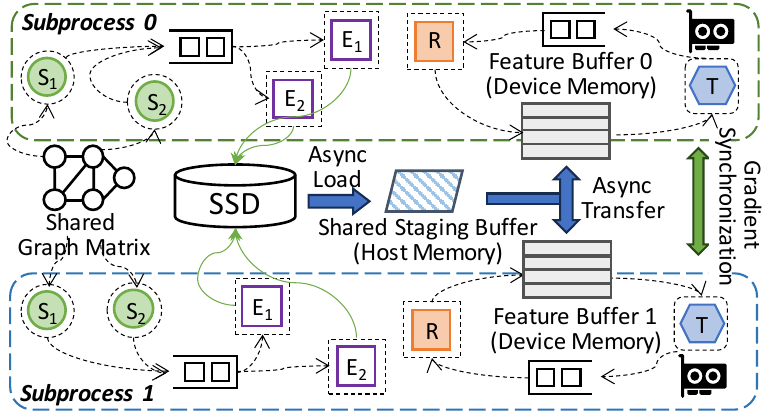}}
	\caption{An illustration of \GNNTrainer's data parallelism.}\label{fig:arch-multi}
\end{figure}

\textbf{Parallelism with Multi-GPUs.}
If multi-GPUs are available,
\GNNTrainer 
 distributes a training workload across them with data parallelism 
 as shown in~\autoref{fig:arch-multi}.  
It utilizes multiple subprocesses, rather than threads,
due to 
Python's Global Interpreter Lock~\cite{py:gil}. 
It divides the entire training set into {\em segments} 
for  subprocesses to execute. 
Each subprocess is responsible for a part of training set (segment) 
with one GPU 
and performs gradient synchronization with other subprocesses being in the backward pass~\cite{train:dataparallel}.
Note that the segment of \GNNTrainer
is different from the {\em partition}s of a graph which MariusGNN uses. 
By its very design, MariusGNN employs one GPU for training and   aims to  train only with in-memory partitions
to avoid I/Os. Swapping partitions is inevitable for MariusGNN at runtime. It also needs to make an orderly sequence
of partitions before training.
Yet \GNNTrainer uses the segments 
 to exploit multi-GPUs for concurrent training. 
At runtime,
\GNNTrainer conducts communications and synchronizations between segments being handled in multi-GPUs.

Though, due to inter-process communication (IPC),
multiprocess training is likely to introduce  additional overhead. 
Each subprocess has its own samplers, extractors, releasers, trainer, and  queues 
to eliminate unnecessary IPC. 
To reduce the memory consumption of a subprocess, 
topological data and staging buffer 
are shared among subprocesses. 
A subprocess has its own feature buffer to separately store node features in its GPU's device memory. 
As to the staging buffer, 
each subprocess reserves a portion 
to avoid contention in sharing.
If a subprocess exhausts its portion, 
it can temporarily 
 ask for extra space from the staging buffer. 
To minimize 
I/Os and improve memory utilization, 
a subprocess is allowed to use data from other subprocesses' portions for the second phase of extraction 
if the data is already present in the shared staging buffer.

\subsection{Implementation and Discussions}

{\bf Implementation.}
We implement  \GNNTrainer with PyG~\cite{gnn:pyg}. 
We utilize the io\_uring \cite{tool:iouring} and CUDA Toolkit \cite{tool:cuda}  for aforementioned 
asynchronous   loading and transfer, respectively.
The io\_uring works well with the direct I/O mode, 
which complements our intention of 
alleviating memory contention.
\GNNTrainer does  memory-mapped sampling  like PyG+.
It utilizes the OS's page cache to hold the index array (e.g., {\tt indices} in SciPy~\cite{tool:scipy}) of
 adjacency matrix 
and keeps the index pointer array (e.g., {\tt indptr} in SciPy) in memory. 
The sampler in \GNNTrainer  supports various 
sampling policies 
and domain-specific node caching methods~\cite{gnn:ginex,gnn:gnnlab,gnn:AliGraph} with high adaptability.

{\bf Access Granularity.} 
Direct I/O  enforces a constraint that
data must be accessed in a multiple of sector size (512B). 
With  typical feature data types, such as float32 (4B), 
a loading I/O request demands a minimum dimension size of 128. 
To facilitate loading with smaller   dimensions, e.g., 32 and 64, 
\GNNTrainer 
combines features belonging to neighboring nodes 
for joint extraction. 
While this may load some data redundantly in the extraction stage,
the overhead is negligible compared to the overall training time (see Section~\ref{sec:eval:overall}).
Moreover, continuous joint exaction 
is likely to exploit spatial locality 
among nodes. 
As to features that do not precisely fit the access granularity, e.g., 127 or 129, 
\GNNTrainer needs to perform extraction with redundant data in the staging buffer, resulting in certain space wastage. 
If the memory space suffices, it can leverage buffered I/Os with asynchronous loading 
to avoid such redundancy.

{\bf GPU Direct Access.} 
GPUs with   
direct memory access enable 
a new I/O path between  SSD  and GPU via PCIe bus~\cite{nvme:gds,nvme:amd,gds:BaM,nvme:cufile}.
\GNNTrainer can employ this technology to eliminate the staging buffer in host memory 
and further reduce memory contention. 
However, GPU direct access currently has few limitations. 
For example, GPUDirect Storage (GDS)  
only supports NVMe SSDs with 
a limited range of GPU products \cite{nvme:gds}.
Worse, as GDS needs
an access granularity of 4KB,
redundant loading is inevitable in the extract stage.
We leave how to incorporate such technologies in \GNNTrainer 
as our future work.

{\bf CPU-based Training.}
\GNNTrainer offers support for CPU-only training architecture~\cite{gnn:bytegnn}. 
It needs to hold 
the feature buffer 
in the host memory. 
Feature data is asynchronously extracted  from SSD and directly sent to the feature buffer, 
without the need of transfer via a staging buffer. 
The management  of   feature buffer remains similar to GPU-based training architecture. 
As to employing data parallelism with multiple subprocesses for CPU-based training, 
\GNNTrainer still shares the feature buffer among subprocesses.

\begin{table}[t]
	\caption{A summary of datasets (M: million; B: billion).}\label{tab:dataset}
	\centering
	\resizebox{\columnwidth}{!}{
		\begin{tabular}{cccccccc}
			\hline
			\multirow{2}{*}{\textbf{Dataset}} & \multirow{2}{*}{\textbf{\#Node}} & \multirow{2}{*}{\textbf{\#Edge}} & \multirow{2}{*}{\textbf{Dim.}} & \multirow{2}{*}{\textbf{\#Class}} & \multicolumn{3}{c}{\textbf{Memory (GB)}} \\ \cline{6-8} 
			&  &  &  &  & Topo. & Feat. & Tol. \\ \hline
			Papers100M~\cite{dataset:ogb} & 111M & 1.6B & 128 & 172 & 13 & 53 & 67 \\
			Twitter~\cite{dataset:twitter} & 41.7M & 1.5B & 128 & 50 & 11 & 20 & 31 \\
			Friendster~\cite{dataset:friendster} & 65.6M & 1.8B & 128 & 50 & 14 & 32 & 46 \\
			MAG240M~\cite{dataset:ogblsc} & 122M & 1.3B & 768 & 153 & 10 & 349 & 359 \\ \hline
	\end{tabular}}
\end{table}

\section{Evaluation}\label{sec:eval}

\begin{figure*}[t]
	\centering
	\includegraphics[width=\linewidth,page=1]{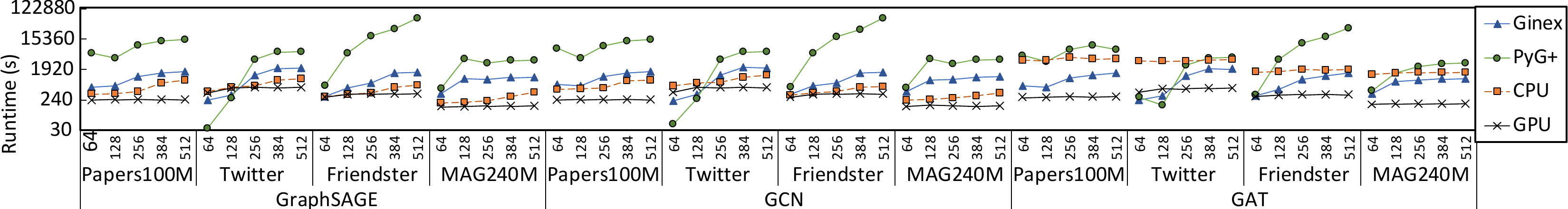}
	\caption{The runtime of one epoch in Ginex, PyG+, and \GNNTrainer with varying feature dimensions. }\label{fig:eval-dim}
\end{figure*}

\begin{figure*}[t]
	\centering
	\includegraphics[width=\linewidth,page=1]{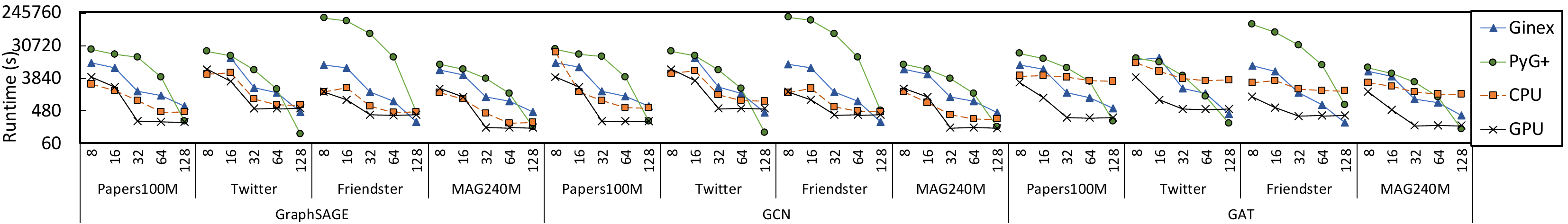}
	\caption{The runtime of one epoch in Ginex, PyG+, and \GNNTrainer with    varying memory sizes.}   \label{fig:eval-mem}
\end{figure*}

\begin{figure*}[t]
    \centering
    \includegraphics[width=\linewidth,page=1]{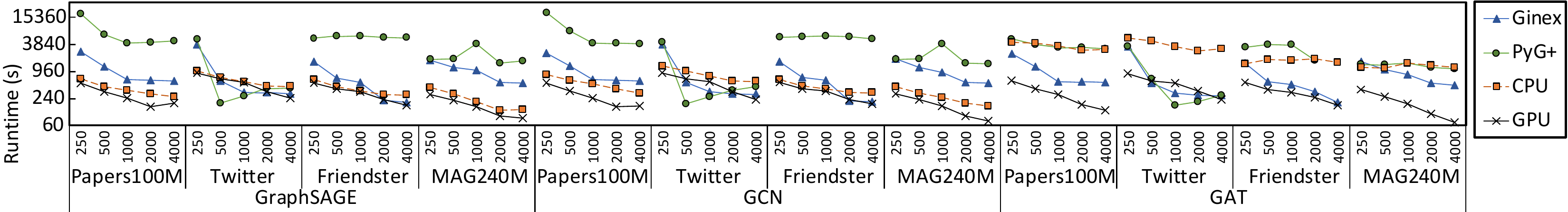}
    \caption{The runtime of one epoch in Ginex, PyG+, and \GNNTrainer with
    varying mini-batch sizes.}    \label{fig:eval-batch}
\end{figure*}

{\bf Platform.} 
We conduct experiments on a machine with
two Intel Xeon Gold 6342 CPUs, 
two NVIDIA GeForce RTX 3090 GPUs (24GB device memory), 
and a SAMSUNG PM883 SSD. 
Regarding the sizes and scales 
of public datasets (see \autoref{tab:dataset}),
we configure the default capacity of host memory as 32GB. 
This suite of hardware devices is ordinarily available in physical or cloud machines.
The OS is Ubuntu 22.04 with Linux kernel 5.19.0. 
Other relevant frameworks include
Python v3.8, PyTorch v1.12, CUDA v11.6, and PyG v2.2.

{\bf GNN Models.}
We  evaluate \GNNTrainer with 
three  GNN models, i.e.,
GraphSAGE~\cite{model:sage}, GCN~\cite{model:gcn}, and GAT~\cite{model:gat}.
All of them are configured with 3 layers, 
3-hop random neighborhood sampling, and a dimension of 256 for hidden layer. 
For GraphSAGE and GCN, we set the sampling size as (10, 10, 10),
and  this parameter is 
(10, 10, 5) for GAT. 
We set
the default mini-batch size 
as 1,000. 
We decide these
settings (e.g., the number of model layers) 
by referring to prior works~\cite{gnn:ginex,gnn:gnnlab}
and respective 
official   guidelines or examples.

{\bf Datasets.}
We take four datasets listed in~\autoref{tab:dataset} for evaluation, i.e.,
Papers100M~\cite{dataset:ogb}, Twitter~\cite{dataset:twitter}, Friendster~\cite{dataset:friendster}, and MAG240M~\cite{dataset:ogblsc}. 
Following the common practice of prior works~\cite{gnn:ginex,gnn:gnnlab,gnn:dorylus},
we generate random features and labels for Twitter and Friendster as they innately lack such information.
By referring to~\cite{gnn:mariusgnn}, we utilize only the paper nodes and citation edges for MAG240M. 
The default feature dimension for all datasets is set as 128. 
The topological data is stored in a compressed sparse column (CSC)-formatted adjacency matrix for each dataset. 
For \GNNTrainer and baselines, 
we keep in memory the index pointer array 
of  adjacency matrix 
since it occupies less than 1GB and is frequently accessed in the sample stage.
This aligns with the setup in~\cite{gnn:ginex}. 
Other data, including the index array 
of adjacency matrix, 
is stored in SSD during training.

\begin{figure}[t]
	\centering
	\begin{subfigure}{0.49\linewidth}
		\includegraphics[width=\linewidth,page=1]{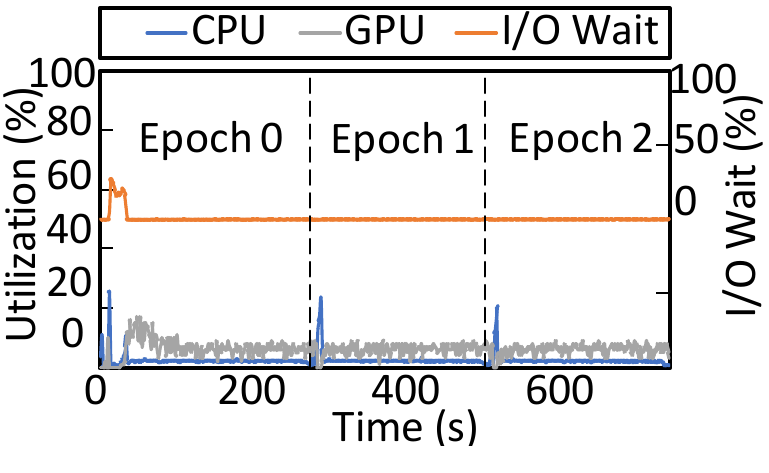}
		\caption{GPU-based training.}
		\label{fig:eval-io:gpu}
	\end{subfigure}	
	\hfill
	\begin{subfigure}{0.49\linewidth}
		\includegraphics[width=\linewidth,page=2]{eval-io.pdf}
		\caption{CPU-based training.}
		\label{fig:eval-io:cpu}
	\end{subfigure}
	\caption{CPU utilization, GPU utilization and percentage of I/O wait time with \GNNTrainer for three epochs.}
	\label{fig:eval-io}
\end{figure}

{\bf Baselines.}
We consider open-source  PyG+, Ginex, and MariusGNN 
 as   baselines. 
In line with~\cite{gnn:ginex,gnn:mariusgnn},
they all use one GPU for training.
Following~\cite{gnn:ginex}, we set
the number of threads used for I/O-intensive operations (e.g., extraction) 
as more than twice the number of physical CPU cores 
to maximally utilize SSD's bandwidth. 
Ginex is equipped with a 6GB neighbor cache and a 24GB feature cache by default. 
The superbatch size of Ginex is set as 1,500. 
As mentioned,
MariusGNN generates a 
sequence of partitions,
while others have no such data preparation.
Moreover, MariusGNN can only finish a part of workloads. 
It encounters  OOM issue for GAT model and  
does not well support GCN model.
We hence individually compare it to \GNNTrainer in Section \ref{sec:mariusgnn}.
We test both GPU- and CPU-based  \GNNTrainer variants, denoted as
{\tt GPU} and {\tt CPU}, respectively, in Figures \ref{fig:mot-sample:dim}, \ref{fig:eval-dim} to \ref{fig:eval-acc}.
We empirically configure \GNNTrainer to have
 four samplers, four extractors, one trainer, and one releaser. 
The capacity bounds of extracting and training queues are six and four, 
which are a bit greater or equal to the numbers of samplers and extractors, respectively.
At runtime
samplers and extractors would  be blocked if corresponding queues are full.

All results are averaged over 10 epochs. The primary metric is the average time per epoch. 
Shorter time means higher performance.

\subsection{Training Performance}\label{sec:eval:overall}

{\bf Overall performance.}
We  compare \GNNTrainer to PyG+ and Ginex with all combinations of graph datasets and training models. 
As shown in
\autoref{fig:eval-dim},
GPU-based \GNNTrainer consistently outperforms all others
on serving those workloads.
Take Papers100M for example.
GPU-based \GNNTrainer makes 16.9$\times$ and 2.6$\times$ speedups than  PyG+ and Ginex, respectively,
when all of them employ 
GraphSAGE and GCN models with a dimension size of 128.
With the GAT model, 
 \GNNTrainer yields 11.2$\times$ 
and 2.0$\times$ speedups, respectively. 

 \GNNTrainer's performance gains are mainly attributed to its reduced memory contention and efficient asynchronous 
extraction. 
PyG+  synchronously loads data from SSD when the required memory-mapped feature data is not in the OS's page cache. 
Memory contention between  sample   and extract stages further degrades the performance of PyG+. 
 Ginex tries to alleviate memory contention through separate  caches, but
is still hindered by the synchronous initialization of feature cache for each superbatch. 
Additionally, Ginex incurs substantial cost due to inspect operations in 
pursuit of an optimized cache replacement policy. 
When doing so, Ginex must sample data and store into SSD the sampling results in advance  per superbatch, 
resulting in extra I/Os and costly inspect operations.

CPU-based \GNNTrainer yields different but generally matching effects with GPU-based variant  for  GraphSAGE and GCN models. With them,
GPU-based variant achieves   1.5$\times$ and 2.1$\times$ speedups over CPU-based variant, respectively. 
However,  the performance gap between them is noteworthy for the GAT model.
For example, with Papers100M,
CPU-based \GNNTrainer is 12.1$\times$ slower than GPU-based \GNNTrainer. 
In some datasets with the GAT model, CPU-based \GNNTrainer achieves similar or  worse performance than PyG+ and Ginex
that are using GPU for training. 
We find that
CPU-based variant with the GAT model spends 8.0$\times$ execution time on average 
 than GPU-based one.
Thus,
the inferior performance of the former is attributed to the high cost of using CPU to train the GAT model.

{\bf Feature Dimensions}.
We vary 
the size of feature dimension from 64 to 512 for all datasets and training models. 
\autoref{fig:eval-dim} illustrates the runtime, 
with the Y-axis presented in the logarithmic scale. 
Similar trends are observed for all models with \GNNTrainer, PyG+, and Ginex. 
As the feature dimension increases, the runtime of each system increases, since 
a larger volume of feature data 
needs to be loaded and extracted from SSD. 
For example, when the feature dimension increases from 64 to 512 
for MAG240M with the GraphSAGE model, 
GPU-based \GNNTrainer experiences 1.1$\times$ increase in 
training time. 
PyG+   is much more sensitive. 
For instance,
with MAG240M and GraphSAGE model, 
PyG+ costs 7.0$\times$ more training time 
when the dimension increases from 64 to 512.
Meanwhile,
with a smaller dimension for Twitter and Friendster datasets, 
PyG+ achieves comparable or even higher performance than \GNNTrainer. 
This is because
 smaller files that hold datasets with lower feature dimensions
lead to a higher likelihood of being buffered
in the OS's page cache which PyG+ exploits for training. 
Likewise, Ginex's feature cache is also likely to hold the entire feature data with lower   dimensions, thereby resulting in improved performance.

{\bf Memory Capacity}.
We vary the memory capacity from 8GB to 128GB for all datasets with a large dimension size of 512. 
For Ginex, we vary sizes of its caches based on the host memory capacity, 
ensuring that its two caches occupy at least 85\
The results are shown in~\autoref{fig:eval-mem} with the Y-axis being in the logarithmic scale.
Ginex suffers from OOM issue and 
fails to train GraphSAGE and GCN models with Twitter dataset at a capacity of 8GB.
In general,  \GNNTrainer, PyG+, and Ginex all demonstrate higher performance with larger host memory, 
which directly mitigates the contention between sample and extract stages. 
As mentioned,
PyG+ heavily relies  on the OS's page cache and shows notable  sensitivity to   memory capacity. 
In some datasets, e.g., Twitter,
PyG+ is comparable or even  faster than \GNNTrainer with  128GB memory,
 because PyG+ embraces the
  sufficient page cache to hold all data for such datasets. 
Ginex also yields improved performance when  the host memory increases,
since its increasing caches are able to buffer more data.

However, 
GPU-based \GNNTrainer consistently outperforms other systems in most scenarios. 
Even with 8GB memory,
GPU-based \GNNTrainer is more performant than, for example, PyG+ with 5.8$\times$ gap. 
This is attributed to \GNNTrainer's strict memory footprint in the extract stage, 
which occupies only essential memory space  for sampling. 
As the host memory   increases beyond 32GB, 
the runtime of \GNNTrainer does not evidently drop. 
This is because 32GB memory is sufficient for \GNNTrainer 
to retain a graph's entire topological data with restricted staging buffer. 
However, CPU-based \GNNTrainer performs similarly or even worse than PyG+ and Ginex 
due to the inefficiency of CPU training with the GAT model.

{\bf Mini-batch Sizes}.
The mini-batch size  affects the convergence time to an expected accuracy~\cite{gnn:gnnlab,gnn:ginex}.
We  
depict its impacts on the  runtime of Ginex, PyG+, and \GNNTrainer   in \autoref{fig:eval-batch}. 
PyG+ encounters OOM issue when the mini-batch size is set to 4,000 for Friendster dataset with GAT model.
While an increasing mini-batch generally reduces the end-to-end time of a training epoch for \GNNTrainer and Ginex, 
the runtime of PyG+ fluctuates in some cases.
For example, when the mini-batch increases from 500 to 1,000, 
PyG+'s runtime becomes 43.2\
The size of mini-batch  affects the size of memory space   PyG+ uses for the feature data
in the extract stage. 
A larger mini-batch 
competes for more memory space demanded by sampling.
As mentioned in Section~\ref{sec:motivation},
the sample stage of PyG+ is highly sensitive to memory contention,
which entails prolonging the sampling time and in turn causes evident   inefficiency.

\subsection{Deep Dissection with \GNNTrainer}\label{sec:eval:deepdive}

{\bf Reduced Memory Footprint.}
To assess if \GNNTrainer  relieves the memory contention, 
we measure the sampling time for both `-only' and `-all' situations
(see Section~\ref{sec:motivation}).
For a back-to-back comparison,
the results are shown in~\autoref{fig:mot-sample:dim} with those of PyG+ and Ginex. 
Firstly, due to the reduced memory footprint used by extractors, 
\GNNTrainer greatly reduces
the sampling time. 
For example, 
compared to PyG+, 
GPU-based \GNNTrainer achieves 1.9$\times$ and 3.0$\times$ speedups with 128  and 512 dimensions, respectively.
Secondly, when the dimension size increases,
\GNNTrainer does not experience severe memory contention.
GPU-based \GNNTrainer spends comparable time on datasets with varying dimensions.
The sampling time of CPU-based \GNNTrainer 
increases marginally with increased dimension sizes. 
Unlike GPU-based training that puts feature data in GPU's device memory,
CPU-based training does so with the host memory.
The latter thus demands more memory space with higher feature dimensions,
which in turn affects the sampling time.

{\bf Reduced I/O Congestion.}
\GNNTrainer asynchronously loads and transfers feature data 
in 20GB to 349GB outside of the critical path during training with four datasets
(see \autoref{tab:dataset}).
Without loss of generality, when training with
GraphSAGE model and Papers100M  for three epochs, we
monitor CPU utilization, GPU utilization, and the ratio of I/O wait time. 
\autoref{fig:eval-io} illustrates the results for both GPU- and CPU-based \GNNTrainer. 
Compared to 
PyG+ and Ginex (see~\autoref{fig:mot-io}), 
\GNNTrainer largely reduces I/O wait time with asynchronous I/Os. 
Additionally, \GNNTrainer dedicates
one extractor thread to solely  handling  
 all asynchronous I/Os performed to extract nodes in two consecutive phases of the extract stage 
for a mini-batch (\mininumbercircled{4} and \mininumbercircled{5}
in \autoref{fig:arch-gpu}).
This 
 improves the CPU utilization, since context switches that frequently happen 
 to  multi-threaded synchronous I/Os are almost eliminated.

 \begin{figure}[t]
	\centering
	\includegraphics[width=\linewidth,page=1]{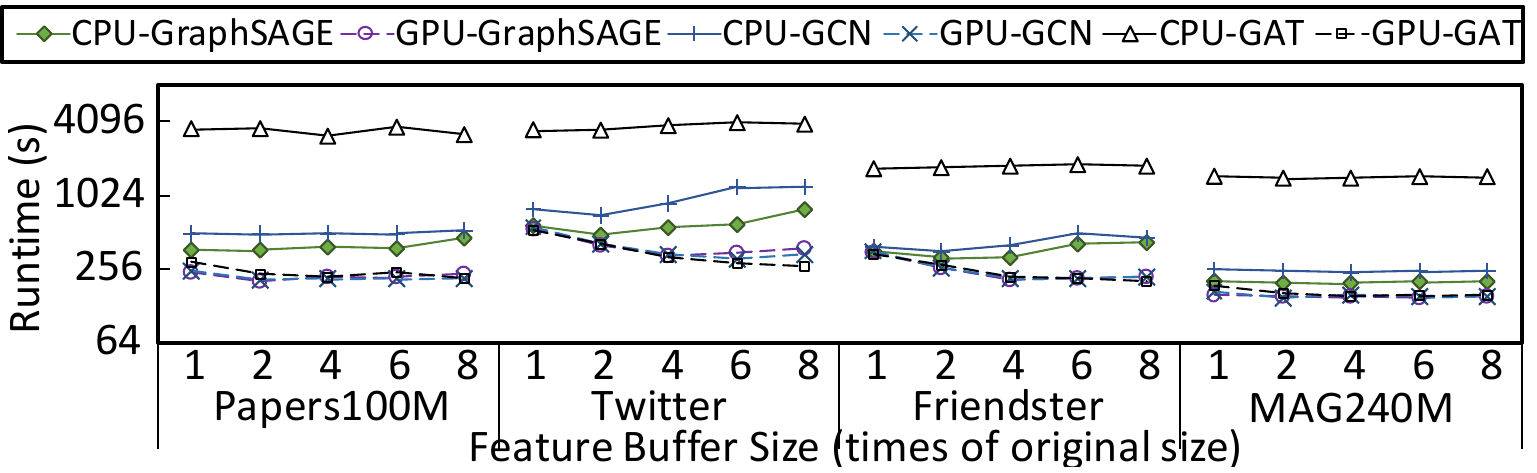}
	\caption{The runtime of one epoch in \GNNTrainer with
		varying feature buffer sizes.}\label{fig:eval-buffer}
\end{figure}

{\bf Feature Buffer Size.}
\autoref{fig:eval-buffer} captures 
the impact of feature buffer size
on \GNNTrainer's performance. 
Specifically, we scale up 
the feature dimension from 2$\times$ to 8$\times$ 
of the default setting (approximately 2.38GB). 
When the feature buffer size doubles (2$\times$), 
\GNNTrainer's performance improves by exploiting data locality between mini-batches. 
For instance, \GNNTrainer 
achieves speedups of 1.4$\times$ and 1.2$\times$ 
over the original feature size for GPU- and CPU-based variants, respectively, 
with Twitter and GraphSAGE model. 
However,  increasing the feature buffer size to be even greater does not  lead to further   improvement.
This is due to the decreasing cost efficiency. For example, a
larger feature buffer incurs more overhead in management, such as updating the standby list.

{\bf Scalability}.
To test if \GNNTrainer  scales with multi-GPUs,
we use another  machine that has eight NVIDIA Tesla K80 GPUs (12GB memory per GPU), 
two Intel Xeon E5-2690 CPUs, 
and an Intel DC S3510 SSD. 
This is an economical  machine, as all hardware devices 
were made 8 to 10 years ago.
The host memory  is not restricted (256GB) 
since 
\GNNTrainer is 
not very sensitive to memory capacity (see Section~\ref{sec:eval:overall}).
We vary the number of subprocesses.
For GPU-based \GNNTrainer, a
 subprocess is exclusively assigned per GPU.

\begin{figure}[t]
	\centering
	\includegraphics[width=\linewidth,page=1]{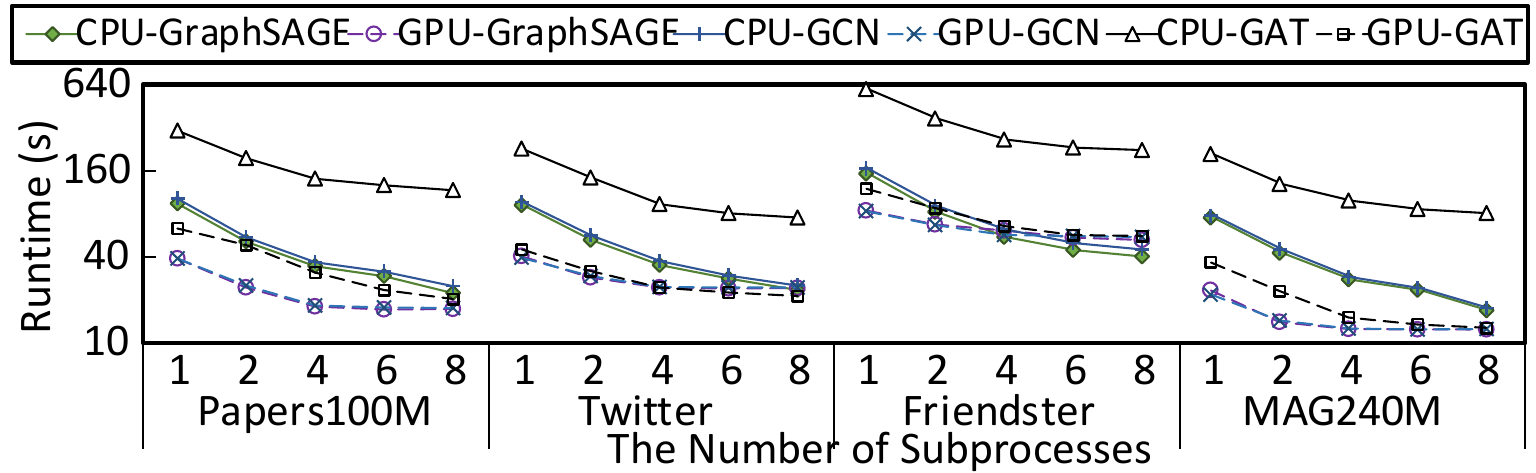}
	\caption{The scalability of \GNNTrainer with Multi-GPUs.}\label{fig:eval-multi}
\end{figure}

As shown in \autoref{fig:eval-multi}, 
the epoch time of both CPU-  and GPU-based {\GNNTrainer} variants initially decreases as the number of subprocesses increases. 
For instance, when handling  MAG240M  with the GraphSAGE model, \GNNTrainer with two subprocesses makes a speedup of 1.7$\times$ and 1.8$\times$ over a single subprocess for GPU-  and CPU-based training, respectively. 
This improvement is because each subprocess has a smaller number of training nodes to process. 
However, due to the IPC cost and gradient synchronization, the setting with
two subprocesses is a bit lower than 2$\times$ speedup. 
Notably, in the case of GPU-based \GNNTrainer, the epoch time no longer decreases further with a larger number of subprocesses (six subprocesses) 
as the synchronization overhead between GPUs turns to be the bottleneck, consuming more GPU resources and PCIe bus bandwidth. 
Furthermore, 
with the increase of  subprocesses, 
the number of epochs required to converge increases. This aligns with the observations of Yang et al.~\cite{gnn:gnnlab}.

\subsection{Training Convergence}\label{sec:eval:convergence}

\begin{figure}[t]
	\centering
	\begin{subfigure}{0.49\linewidth}
		\includegraphics[width=\linewidth,page=1]{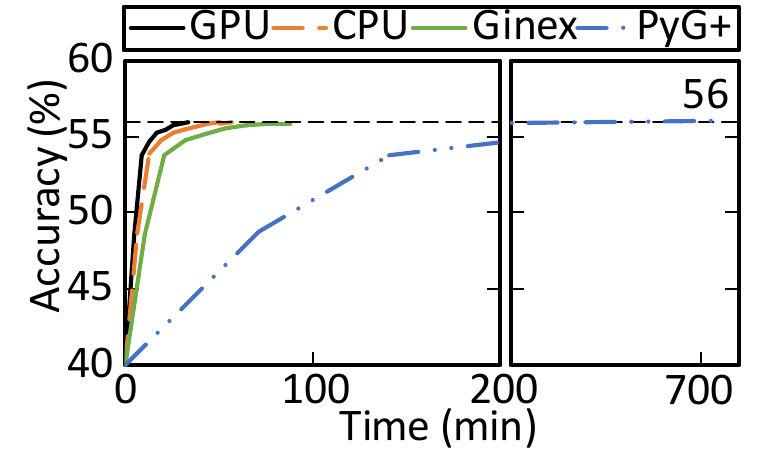}
		\caption{Papars100M.}
		\label{fig:eval-acc:paper}
	\end{subfigure}	
	\hfill
	\begin{subfigure}{0.49\linewidth}
		\includegraphics[width=\linewidth,page=2]{eval-acc.pdf}
		\caption{MAG240M.}
		\label{fig:eval-acc:mag}
	\end{subfigure}
	\caption{A comparison of time-to-accuracy using Ginex, PyG+, and \GNNTrainer for training GraphSAGE model.}\label{fig:eval-acc}
\end{figure}

\GNNTrainer makes mini-batch reordering with multiple samplers and extractors,
which is not covered by   Ginex and PyG+.
Thus, it is necessary to evaluate the impact of mini-batch reordering 
and verify the correctness of \GNNTrainer. 
Figures~\ref{fig:eval-acc:paper} and~\ref{fig:eval-acc:mag} present 
the time-to-accuracy curves of different training frameworks 
for Papers100M with a dimension  size of 128 and MAG240M with a dimension  size of 768, respectively, using the GraphSAGE model. 
The mini-batch reordering does not affect convergence.
\GNNTrainer converges in similar or  fewer epochs than PyG+. 
Take   Papers100M   for illustration. 
All   systems converge to the same accuracy target (about 56\
PyG+, Ginex, and CPU-based \GNNTrainer cost 18.4$\times$, 2.9$\times$, and 1.6$\times$ runtime, respectively, than GPU-based \GNNTrainer.
Both   GPU-  and CPU-based {\GNNTrainer} variants  benefit from efficient asynchronous extraction and mini-batch reordering. 
As to MAG240M, only GPU-based \GNNTrainer reaches the target accuracy (about 52\
while PyG+ and Ginex encounter OOT (out of time) and OOM issues, respectively.
This affirms the robustness of \GNNTrainer.

\subsection{A Comparison with MariusGNN}\label{sec:mariusgnn}

We evaluate MariusGNN against \GNNTrainer 
using Papers100M (128 dimensions) and MAG240M (768 dimensions)   with the GraphSAGE model.
As told by preceding tests,
these two datasets are constrained by 32GB host memory.
 \autoref{tab:marius} compares the runtime of an epoch.
 Note that 
{\it Data Preparation} 
refers to the time taken by MariusGNN only 
to order a sequence of partitions and preload graph data. 

From~\autoref{tab:marius}, we 
obtain following   observations. 
Firstly, 
  GPU-based \GNNTrainer still demonstrates superior performance over MariusGNN, 
achieving a \newfpeval{round(346.66/241.12, 1)}$\times$ speedup of {\it training time}
with Papers100M.
This is because, although MariusGNN tries to avoid I/Os during training,
it still suffers from longer I/O wait time at the beginning of training (see~\autoref{fig:mot-io:marius-32}). 
Secondly, MariusGNN severely suffers from mandatory and inefficient data preparations, 
which lead to \newfpeval{round(643.02/241.12, 1)}$\times$ {\it overall time} than GPU-based \GNNTrainer.
Thirdly, MariusGNN fails to effectively handle larger graphs. 
For   MAG240M   with a feature dimension of 768, 
MariusGNN cannot finish training in 32GB memory due to
OOM issue. 
This implies the limited  feasibility of MariusGNN. 
We have done an additional test by increasing the host memory to be 128GB for MariusGNN.
As shown by the bottom row of \autoref{tab:marius},
it still cannot complete training with MAG240M because OOM issue has happened during data preparation.
This comparatively justifies the robustness and viability of \GNNTrainer that manages to finish training
MAG240M even with 8GB memory (see Section~\ref{sec:eval:overall}).
For Papars100M,
MariusGNN spends \newfpeval{round(292.17/241.12, 1)}$\times$ overall time than GPU-based \GNNTrainer. 
The cost of data preparation remains high even with sufficient 128GB memory,
taking \newfpeval{round(115.03/292.17*100, 1)}\
To sum up, 
\GNNTrainer not only outperforms SoTA disk-based GNN training systems  
but also shows higher usability. 

\begin{table}[b]
    \caption{The runtime of one epoch in MariusGNN and \GNNTrainer (N/A: not applicable; OOM: out-of-memory.)}\label{tab:marius} 
    \resizebox{\columnwidth}{!}{
    \begin{tabular}{crrrrrr}
    \hline
    Runtime (s)       & \multicolumn{2}{c}{Data Preparation} & \multicolumn{2}{c}{Training} & \multicolumn{2}{c}{Overall} \\ \hline
    Dataset &
      \multicolumn{1}{c}{\begin{tabular}[c]{@{}c@{}}Papers-\\100M\end{tabular}} & 
      \multicolumn{1}{c}{\begin{tabular}[c]{@{}c@{}}MAG-\\240M\end{tabular}} &
      \multicolumn{1}{c}{\begin{tabular}[c]{@{}c@{}}Papers-\\100M\end{tabular}} &
      \multicolumn{1}{c}{\begin{tabular}[c]{@{}c@{}}MAG-\\240M\end{tabular}} &
      \multicolumn{1}{c}{\begin{tabular}[c]{@{}c@{}}Papers-\\100M\end{tabular}} &
      \multicolumn{1}{c}{\begin{tabular}[c]{@{}c@{}}MAG-\\240M\end{tabular}} \\ \hline
      \GNNTrainer-GPU   & N/A          & N/A         & 241.12        & 166.66       & 241.12      & 166.66      \\
      \GNNTrainer-CPU   & N/A          & N/A         & 371.69        & 716.98       & 371.69      & 716.98      \\
      PyG+  & N/A          & N/A         & 4,115.73    & 5,037.61    & 4,115.73      &  5,037.61   \\
      Ginex & N/A          & N/A         & 636.89    & 1,271.55       & 636.89      & 1,271.55      \\
      MariusGNN-32G  & 296.35        & OOM          & 346.66        & OOM          & 643.02      & OOM         \\
      MariusGNN-128G & 115.03        & OOM       & 177.14        & OOM       &  292.17     & OOM      \\ \hline
    \end{tabular}
    }
\end{table}
\section{Related Work}\label{sec:related}

{\bf Disk-based GNN Training.}
Researchers have taken a few approaches to accelerate GNN training with large 
graphs.
\GNNTrainer falls into the category of disk-based training, the objective of which is to overcome the memory capacity limitation when processing massive graph datasets~\cite{gnn:mariusgnn,gnn:ginex,gnn:SmartSAGE}.  
We have  studied Ginex, PyG+, and MariusGNN.  \GNNTrainer outperforms them as it not only deals with memory contention, but also takes into account other limitations such as I/O congestion that commonly arises on ordinary machines.
Besides GNN,
I/O is also a critical issue for other large-scale graph workloads~\cite{graph:GraphWalker,graph:RealGraph}.
For instance,
GraphWalker~\cite{graph:GraphWalker} deploys a state-aware I/O model with asynchronous walk updating 
and utilizes asynchronous batched I/Os to write back walk states  to storage.
\GNNTrainer has some similarities with these graph processing systems but
encounters specific challenges in training GNNs.

{\bf Sample-based GNN Training.} 
Sampling enables efficient neighborhood feature aggregation 
using dense tensor operations. 
Many SoTA GNN training systems execute the train stage on GPUs.
Some of them 
primarily perform the sample and extract stages on CPUs~\cite{gnn:AliGraph,gnn:bytegnn},
while others utilize GPUs for node sampling~\cite{gnn:gnnlab,gnn:LazyGCN}. 
Comparatively,
\GNNTrainer focuses 
on the memory utilization and I/O operation in sample and extract stages,
which allows it to  support various sampling methods and policies.

{\bf Distributed GNN Training.}
Distributed   training partitions a graph dataset into subgraphs and handles them with a cluster of machines~\cite{gnn:AliGraph,gnn:bytegnn}.
For example,
AliGraph~\cite{gnn:AliGraph} adopts sampling-based distributed GNN training and 
caches nodes on local machines to reduce network communications.   
In order to efficiently access graph topological and feature data,
DistDGL~\cite{gnn:DistDGL} employs a distributed in-memory key-value store. 
ByteGNN~\cite{gnn:bytegnn} does CPU-based distributed GNN training. 
It 
has comprehensive scheduling and partitioning algorithms among multiple machines
to improve resource utilization and reduce  training time.
Compared to \GNNTrainer running on a local ordinary machine,
these systems demand multiple machines 
with much higher monetary cost.

{\bf Whole-graph GNN Training.}
Whole-graph training  divides a large graph into  partitions 
and trains GNN models on all nodes or edges simultaneously using multiple machines or multi-GPUs of a machine~\cite{gnn:dorylus,gnn:ROC,gnn:hongtu:sigmod23}.
One  challenge of doing it is that, during 
the aggregation of neighborhood features, 
each node has to consider all neighbors that may have various sizes. 
Hence it 
is likely to severely suffer from memory contention, I/O congestion, and furthermore issues in the training procedure.
We plan to study and 
  accelerate  whole-graph  training in the near future.

\section{Conclusion}\label{sec:conclusion}

Disk-based GNN training systems suffer from performance penalties caused by memory contention and I/O congestion.
We accordingly develop \GNNTrainer. 
\GNNTrainer employs a systematic inter-stage buffer management that more supports the sample stage 
 to relieve memory contention.
In the extract stage, it schedules I/O and memory operations to be asynchronous in order to avoid I/O congestion on the critical path. 
\GNNTrainer also makes the most of software  and hardware for further efficient training. 
Experiments confirm that it largely outperforms SoTA GNN training systems.
GNNs exhibit high usefulness and efficiency  in production environments. 
Regarding the affordability and viability of ordinary physical or cloud machines, 
\GNNTrainer provides a promising 
 solution 
for small and medium enterprises as well as academic scientists
to train GNNs.

\begin{acks}
  We sincerely thank the reviewers and  TPC of the 53rd International Conference on Parallel Processing (ICPP '24) for their valuable comments and suggestions.
  We are grateful to Dr. Cheng Chen for an early discussion. 
  We also express sincere gratitude to Mr. Tianming Wen and Mr. Qun Xu for their support of a machine with multi-GPUs for us to do comprehensive experiments.
  This work was supported by
Natural Science Foundation of Shanghai under Grants No. 23ZR1442300 and 22ZR1442000, and ShanghaiTech Startup Funding.
\end{acks}

\bibliographystyle{ACM-Reference-Format}
\bibliography{sample}

\appendix
\section*{Appendix}
\renewcommand\thefigure{\thesection.\arabic{figure}}    
\renewcommand\thetable{\thesection.\arabic{table}} 
 
\setcounter{figure}{0} 
\setcounter{table}{0} 

In Sections \ref{app:async-intro} and \ref{app:async-compare}, we discuss the effect of asynchronous I/Os by a comparison
to synchronous I/Os. We also present a detailed algorithm to help
readers follow main steps of \GNNTrainer's asynchronous extraction (Section \ref{app:algorithm}).

\section{Asynchronous I/O}\label{app:async-intro}
Asynchronous I/O plays a vital role in efficiently managing massive I/O requests
~\cite{async:ibm,async:rocksdb,graph:Graphene,graph:GraphWalker,graph:RealGraph}.
By processing I/O operations within a single thread, 
asynchronous I/O eliminates the overhead of OS context switches, 
which is particularly useful for extensive scaling and concurrency control.
Without loss of generality, let us take io\_uring~\cite{tool:iouring} for illustration.
The io\_uring  is an asynchronous I/O framework provided by Linux kernel. 
It utilizes ring buffers as the primary interface for communication between kernel and user spaces, 
reducing the overhead associated with system calls and data copy. 
The io\_uring employs two ring buffers, i.e., a submission queue (SQ) 
and a 
completion queue (CQ) for submitting a request and receiving completion signal, respectively.
I/O requests are rephrased as submission queue entries (SQEs) and added to the SQ. 
The kernel processes the submitted requests and appends completion queue events (CQEs) to the CQ. 
Users read the CQEs from the head of CQ to obtain the status of each request. 
Overall, io\_uring provides an efficient and high-performance solution for asynchronous I/O programming, offering advantages such as low overhead, reduced system calls, and the potential for zero-copy data transfer~\cite{tool:iouring}.

\section{Comparison between Synchronous and Asynchronous I/Os}\label{app:async-compare}

\begin{figure}[t]
	\centering
	\begin{subfigure}{0.485\columnwidth} 
		\includegraphics[width=\columnwidth,page=3]{mot-async2.pdf}
		\caption{Bandwidth for sync I/O with different threads.}
		\label{fig:mot-async:sync-ban}
	\end{subfigure}	
	\hfill
	\begin{subfigure} {0.485\columnwidth} 
		\includegraphics[width=\columnwidth,page=4]{mot-async2.pdf}
		\caption{Bandwidth for async I/O with different I/O depths.}
		\label{fig:mot-async:async-ban}
	\end{subfigure}
	\begin{subfigure}{0.485\columnwidth} 
		\includegraphics[width=\columnwidth,page=1]{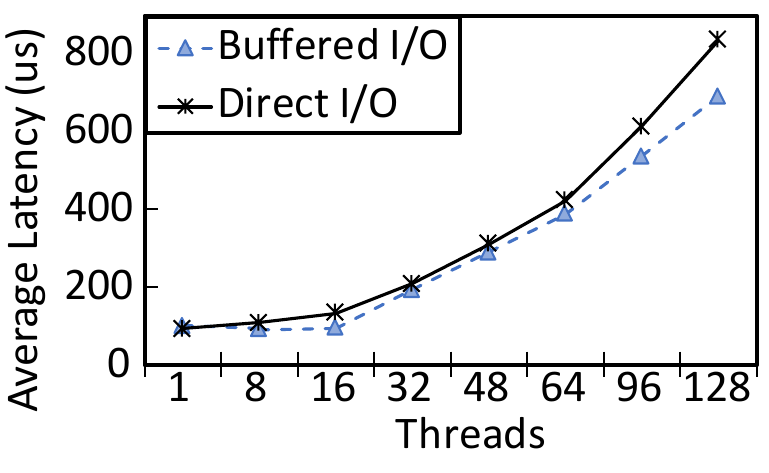}
		\caption{Average latency for sync I/O with different threads.}
		\label{fig:mot-async:sync-lan}
	\end{subfigure}	
	\hfill 
	\begin{subfigure}{0.485\columnwidth} 
		\includegraphics[width=\columnwidth,page=2]{mot-async2.pdf}
		\caption{Average latency for async I/O with different I/O depths.}
		\label{fig:mot-async:async-lan}
	\end{subfigure}
	\caption{A comparison on synchronous I/O with multiple threads and asynchronous I/O with one thread in SSD.}	\label{fig:mot-async}
\end{figure}

\begin{algorithm}[t]	
	\caption{\GNNTrainer's Asynchronous Extraction Procedure}\label{algo:extract}
	\begin{algorithmic}[1]
		\Require The extracting queue $q_e$.
		\State $ids_s$ := {dequeue}($q_e$);
        \Comment{Get sampled node list for a mini-batch}\label{algo:line:ids}
        \State create node alias list $ids_r$ with the same size as $ids_s$; 
        \State set all elements in $ids_r$ to $-1$;\label{algo:line:set-1}
        \State $wait\_list$ := {std::set}();
        \For{($i$ := 0; $i$ < $ids_s$.size(); $i$++)}
        \Comment{Reuse data in feature buffer}
            \State $id$ := $ids_s[i]$; 
            \Comment{Get sampled node ID $id$}
            \If{($map\_table$[$id$].valid == 1)}\label{algo:line:valid}
            \State
            \Comment{The data is already in feature buffer}
                \If{($map\_table$[$id$].ref\_count == 0)} 
                    \State {remove\_slot\_from\_standby\_list}
                    ($map\_table$[$id$].index);\label{algo:line:reuse}
                \EndIf
                \State $ids_r[i]$ := $map\_table$[$id$].index;\label{algo:line:setremap}
            \ElsIf{($map\_table$[$id$].ref\_count > 0)}\label{algo:line:invalid}
                \State
                \Comment{The data is being loaded by another thread}
                \State $wait\_list$.insert($id$);\label{algo:line:insert-wait}
                \State $ids_r[i]$ := $map\_table$[$id$].index;
            \EndIf
            \State $map\_table$[$id$].ref\_count++;\label{algo:line:refp}
        \EndFor
        \For{($i$ := 0; $i$ < $ids_s$.size(); $i$++)}
        \Comment{Load data from SSD}\label{algo:line:loadssd}
            \State $id$ := $ids_s[i]$; 
            \If{($ids_r[i]$ >= 0)}
                \State continue;
                \Comment{Skip data already in feature buffer}\label{algo:line:skip}
            \EndIf
            \State $index$ := {get\_standby\_index}();
            \Comment{Get the LRU standby slot}\label{algo:line:getfree}
            \State $map\_table$[$id$].index = $index$;\label{algo:line:updateindex}            
            \State $reverse\_mapping$[$index$] = $id$; \Comment{Reverse mapping table}
            \State $ids_r[i]$ := $index$;\label{algo:line:updateremap}    
            \State {async\_load\_data\_from\_SSD}($id$);
            \Comment{The first phase}\label{algo:line:phase1}
        \EndFor
        \Repeat \Comment{Transfer data to device memory}\label{algo:line:wait-ssd}
            \State $id$ := {dequeue}($CQ$);
            \Comment{Get node from completion queue}\label{algo:line:deq-cqe}
            \State {async\_transfer\_data\_to\_device}($id$);
            \Comment{The second phase}\label{algo:line:phase2}
        \Until{(all loading tasks are finished)}
        \State wait for the completion of transferring;\label{algo:line:wait-trans}
        \State set $map\_table$[$id$].valid to $1$ for each transferred node $id$;\label{algo:line:setvalid}
        \State wait for the completion of nodes in $wait\_list$;\label{algo:line:wait-wait}
        \State \Return $ids_r$;
	\end{algorithmic}

\end{algorithm}

\textbf{In contrast to synchronous I/O, 
	asynchronous I/O is more suitable for loading massive data 
	and improving CPU utilization as the latter 
	reduces I/O wait time and minimizes context switches.}
We do a standalone I/O test with 
Fio~\cite{benchmark:fio} to compare the effects between synchronous
and asynchronous I/Os.
We randomly read data of a 30GB file in buffered and direct I/O modes, respectively,
and measure the average latency and bandwidth.
The asynchronous I/O is supported by io\_uring~\cite{tool:iouring}.
Each read request aligns with the legacy size of disk sector, i.e., 512B. 
~\autoref{fig:mot-async:sync-ban} and~\autoref{fig:mot-async:sync-lan} 
present the bandwidth and average latency for synchronous I/Os 
(i.e., {\tt read} system call), respectively, with varying thread counts.
~\autoref{fig:mot-async:async-ban} and~\autoref{fig:mot-async:async-lan} 
illustrate the results for asynchronous I/Os issued
with different I/O depths through one thread. 
The I/O depth means in-flight   I/O requests to be served by io\_uring, 
which thus
corresponds to the
queuing depth managed by io\_uring for handling asynchronous I/Os.

{\em For synchronous reading, increasing the number of threads  improves 
the bandwidth in both modes with fewer threads, but soon becomes saturated with increasingly more threads},
typically beyond 32, due to the SSD's hardware limitations.
Worse, multi-threaded reading also results in longer average latencies as depicted in~\autoref{fig:mot-async:sync-lan}. 
This is because of
thread contention~\cite{tool:iouring} and I/O dispatch~\cite{210532}.

More important, 
{\em compared to multi-threading synchronous reading, asynchronous reading achieves a similar effect 
with just a single thread but a greater I/O depth.} 
As depicted in~\autoref{fig:mot-async:async-ban}, the
bandwidth increases as the I/O depth grows. 
Yet the latency also increases due to I/O dispatch,
which emulates the longer time in consecutively extracting data from SSD for disk-based GNN training.
The key advantage of asynchronous reading is 
its ability to avoid 
the OS's context switches 
by processing the most I/Os in one thread. 
As a result, significant I/O wait time is reduced
and 
CPU is relieved from staying idle to wait for the completion of synchronous I/Os. 
These justify the viability of asynchronous I/O   to 
handle massive I/Os for disk-based GNN training.

Last but not the least, 
{\em  buffered I/Os do  not exhibit superb performance compared to direct I/Os,
with regard to both synchronous  
and asynchronous I/Os upon multi-threads and large I/O depths.}
Without loss of generality, let us take asynchronous I/O for analysis. 
The bandwidth of 
buffered I/Os is 41.8\
but the difference narrows down to just 5.6\
buffered I/Os consume the OS's page cache and, if used for GNN training, are likely to
worsen  the aforementioned memory contention. 
Hence, it is practically feasible to use direct I/Os instead of buffered I/Os 
for disk-based GNN training.

\section{Algorithm in \GNNTrainer}\label{app:algorithm}

Algorithm~\ref{algo:extract} illustrates how one extractor of \GNNTrainer asynchronously extracts feature data  from SSD for a training mini-batch. It complements the descriptions shown in Section~\ref{sec:async-extract}. In implementing the function of asynchronous extraction,
we mainly follow structures and routines of io\_uring library, because 
io\_uring
is supported within Linux kernel. 
\GNNTrainer is also implementable with other libraries, such as Asio\footnote{Asio C++ library: \url{https://think-async.com/Asio/}}.

\end{document}